\documentclass[reprint, superscriptaddress, amsmath, amssymb, aps]{revtex4-2}

\usepackage{graphicx}
\usepackage{dcolumn}
\usepackage{upgreek}
\usepackage{xr-hyper}
\usepackage{units}
\usepackage[hidelinks]{hyperref}
\usepackage{braket}
\usepackage{bm}

\begin{document}

\title{Mapping single electron spins with magnetic tomography}

\author{Dan Yudilevich}
\affiliation{
 Department of Chemical and Biological Physics \\ Weizmann Institute of Science, Rehovot 7610001, Israel
}

\author{Rainer St{\"o}hr}
\affiliation{3.\,Physikalisches Institut, Universit{\"a}t Stuttgart, Stuttgart 70569, Germany}

\author{Andrej Denisenko}
\affiliation{3.\,Physikalisches Institut, Universit{\"a}t Stuttgart, Stuttgart 70569, Germany} 

\author{Amit Finkler}
 \email{amit.finkler@weizmann.ac.il}
\affiliation{
 Department of Chemical and Biological Physics \\ Weizmann Institute of Science, Rehovot 7610001, Israel
}

\date{\today}

\begin{abstract}
Mapping the positions of single electron spins is a highly desired capability for applications such as nanoscale magnetic resonance imaging and quantum network characterization. Here, we demonstrate a method based on rotating an external magnetic field to identify the precise location of single electron spins in the vicinity of a quantum spin sensor. We use a nitrogen-vacancy center in diamond as a quantum sensor and modulate the dipolar coupling to a proximate electron spin in the crystal by varying the magnetic field vector. The modulation of the dipolar coupling contains information on the coordinates of the spin, from which we extract its position with an uncertainty of 0.9\,\AA. We show that the method can be used to locate electron spins with nanometer precision up to 10\,nm away from the sensor. We discuss the method's applicability to mapping hyperfine coupled electron spins, and show it may be applied to locating nitroxide radicals. The magnetic tomography method can be utilized for distance measurements for studying the structure of individual molecules.
\end{abstract}

\maketitle

\section{Introduction}
Magnetic resonance spectroscopy (MRS) has been indispensable for determining the structure and function of biomolecules, such as proteins \cite{Schiemann2007, Marion2013}. Electron paramagnetic resonance (EPR), for example, is used to study the structure of organic molecules by measuring the distance between two radicals with unpaired electrons attached to predetermined parts of the molecule (i.e., spin-labels)~\cite{Jeschke2012, Hubbell2013TechnologicalProteins}. Conventional magnetic resonance techniques rely on the signal from large ensembles of molecules and thus measure a mean value. Nanoscale techniques that are sensitive to specific ensemble constituents may augment ensemble techniques, and reveal new information.

The nitrogen-vacancy (NV) center in diamond is an atomic defect in the diamond crystal that can function as a quantum sensor of magnetic fields in nanoscale volumes~\cite{Rondin2012NanoscaleMagnetometer}. Quantum sensing with the NV center is a promising platform for nanoscale MRS~\cite{Cai2013a}, potentially extending the methods down to the single-molecule limit. In recent years, nanoscale nuclear magnetic resonance using NV centers has been demonstrated~\cite{Staudacher2013, Mamin2013a}, down to the single protein level~\cite{Lovchinsky2016}, as well as NV-based EPR spectroscopy of single molecules~\cite{Shi2015a, Schlipf2017}.

By mapping the precise positions of spin-labels attached to individual organic molecules, it would be possible to elucidate the structure of a single molecule. Mapping the positions of individual electron spins is likewise relevant for characterizing organic quantum networks, a proposed platform for quantum processors~\cite{Wasielewski2020}. Such mapping has been demonstrated with magnetic resonance force microscopy~\cite{Mamin2004a}, scanning tunneling microscopy~\cite{Durkan2002ElectronicResonance, Willke2019}, and an NV-based magnetometer coupled with a scanning magnetic tip~\cite{Grinolds2014}.

Mapping the positions of nuclear spins is a related endeavor. Recent studies mapped~$\mathrm{^{13} C}$ spin clusters in a diamond lattice around an NV center sensor based on detecting the spins' Larmor precession\,\cite{Abobeih2019b, Zopes2018b, Cujia2021}. But it is usually impractical for electron spins due to the similarity between the sensor and target spin's gyromagnetic factor.

\begin{figure}
\includegraphics[scale=0.43]{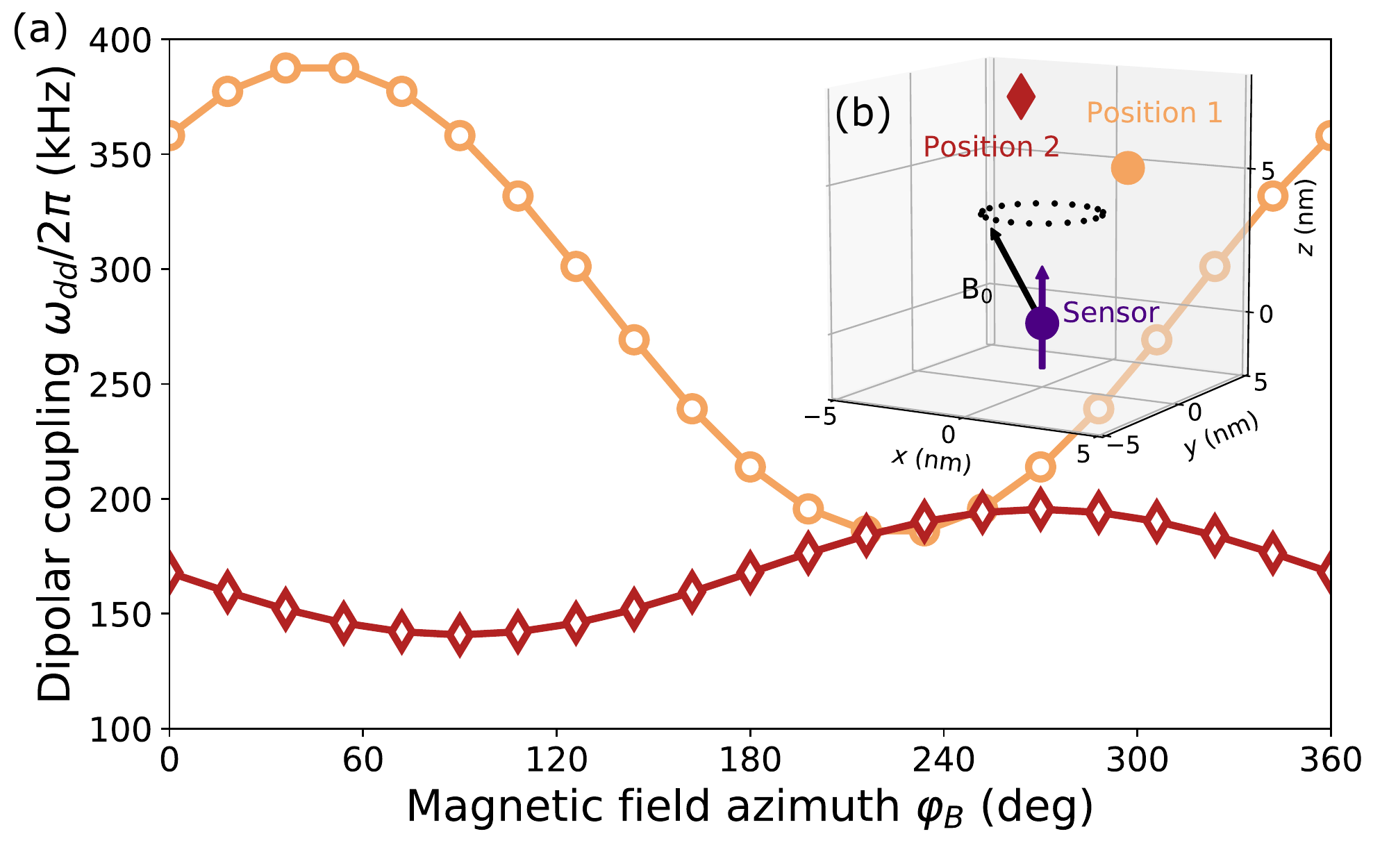}
\caption{\label{fig:two_spins_example} Simulation of the dipolar coupling frequency modulation by a tilted magnetic field. The curves show distinct dipolar coupling oscillations for two target spin positions. The inset depicts the positions relative to the sensor. The simulation was done for a magnetic field tilted at a polar angle $\theta_B=29^{\circ}$ and a varying azimuth $\varphi_B$. The amplitude, offset, and phase of the oscillation contain information on the target spin position.}
\end{figure}

In this Letter, we discuss a protocol for locating electron spins in the vicinity of a quantum spin sensor using a varying magnetic field vector, which modulates the dipolar coupling frequency between a target spin and the sensor spin (see Fig.\,\ref{fig:two_spins_example}). The concept was proposed for mapping nuclear spins~\cite{Cai2013a}, used to map quantum reporter spins on the surface of a diamond for proton magnetic resonance~\cite{Sushkov2014} and locating electron-nuclear spin defects within a diamond~\cite{Cooper2020IdentificationDiamond}. Here, we demonstrate that the method can locate electron spins with {\AA}ngstrom-scale precision, a ten-fold improvement in accuracy, and discuss its applicability for electron spin mapping.

\section{Mapping protocol}

In our protocol, we measure the dipolar coupling $\omega_{dd}$ between the sensor spin and a target spin as a function of the external magnetic field's ($B_0$) orientation. To do so, we tilt the magnetic field from the sensor's axis, and sweep along a full~$\mathrm{360^{\circ}}$ trajectory around the axis, analogous to computerized tomography~\cite{Kak1979ComputerizedSources}. The orientation of $B_0$ modulates the dipolar coupling, such that the position of the spin is encoded in the modulation.

We consider a system of a sensor spin ($\mathbf{S}_{nv}$) with an axial symmetric zero-field splitting ($D$), coupled to a proximate target spin ($\mathbf{S}_{e}$), under an external magnetic field ($\mathbf{B}=B_0 \mathbf{\hat{b}}$). The NV center in diamond is the archetypal sensor; however, other candidate solid-state spin defects may meet these criteria, e.g., SiV in SiC \cite{Nagy2019High-fidelityCarbide}.

For sufficiently distant spins ($\mathrm{\gtrsim 1.5\,nm}$)\cite{Jeschke2002DistanceEPR}, we can approximate the sensor-target spin interaction to a dipole-dipole interaction, denoted by $\mathcal{H}_{dd}$. We choose our coordinate system such that $\mathbf{\hat{z}}$ is the zero-field splitting axis; $\gamma_{nv}$ ($\gamma_{e}$) is the sensor (target) spin gyromagnetic ratio. The sensor-target spin system is thus described by the following Hamiltonian:

\begin{eqnarray}
\mathcal{H}/\hbar &=& D {S_{nv}^{z}}^2 + \gamma_{nv} \mathbf{B} \cdot \mathbf{S}_{nv} + \gamma_e \mathbf{B} \cdot \mathbf{S}_{e} + \mathcal{H}_{dd} / \hbar
\label{eq:spin_hamiltonian}
\end{eqnarray}

We consider a regime where $\mathcal{H}_{dd} \ll \gamma_{nv/e} B_0 \ll D $, so that the eigenstates of the sensor spin are dominated by the zero-field term, and the eigenstates of the target spin are dominated by its Zeeman term. We invoke the secular approximation, and neglect the components of the sensor and target spin operator that do not commute with  $S_{nv}^{z}=\mathbf{S}_{nv}\cdot\mathbf{\hat{z}}$, $S_{e}^{b}=\mathbf{S}_{e}\cdot\mathbf{\hat{b}}$, accordingly. We obtain the following approximated term for the dipole-dipole interaction:

\begin{eqnarray}
\mathcal{H}_{dd} / \hbar & \approx & -\frac{\mu_0 \gamma_{nv} \gamma_{e} \hbar}{2 r^3} \left(3 \left(\mathbf{\hat{z} \cdot \hat{r}} \right) \left(\mathbf{\hat{b} \cdot \hat{r}} \right) - \mathbf{\hat{z} \cdot \hat{r}}  \right) S^{z}_{nv} S_{e}^{b} \nonumber \\
& \equiv & \omega_{dd}(\mathbf{r}, \mathbf{\hat{b}}) S^{z}_{nv} S_{e}^{b}
\end{eqnarray}
\noindent
$ \mathbf{r} = r\mathbf{\hat{r}} $ is the vector connecting the two spins, and we defined $ \omega_{dd}(\mathbf{r}, \mathbf{\hat{b}}) $, the field dependent dipolar coupling strength.

It is convenient to analyze the system in spherical coordinates, where the sensor's position is set as the origin, and the target position is given by the distance $r$, a polar angle $\theta$, and azimuth $\varphi$. We describe the magnetic field orientation by the polar angle $\theta_B$ (the tilt from $\mathbf{\hat{z}}$), and the azimuth $\varphi_B$.
We then write the dipolar coupling strength as a function of the target spin coordinates and the magnetic field orientation:

\begin{eqnarray}
\omega_{dd}(\mathbf{r}, \mathbf{\hat{b}})&=&
-\frac{\mu_0 \gamma_{nv} \gamma_{e} \hbar}{2 r^3} \bigg[\left(3\cos^2\theta-1\right)\cos\theta_B+ \nonumber\\
&+& \frac{3}{2}\sin\theta_B \sin2\theta \cos\left(\varphi_B-\varphi\right) \bigg]
\label{wdd}
\end{eqnarray}

For the case of $ \mathbf{B} \parallel \mathbf{\hat{z}} $, a field aligned with the sensor's axis, the second term of Eq.\,\ref{wdd} vanishes, and the dipolar coupling becomes a function of $r$ and~$\theta$ alone. However, sampling $ \omega_{dd} $ at several magnetic field orientations (i.e., several sets of $ (\theta_B,\,\varphi_B) $) provides information to identify the target spin position.

To extract the available information on the position of the target spin from the dipolar coupling, we consider an experiment where we vary the magnetic field orientation to extract $ (r,\,\theta,\,\varphi)$. Eq.\,\ref{wdd} has the form of a shifted sine, so it is convenient to perform a tomography-like sweep of the magnetic field's azimuth $\varphi_B$ at a constant tilt angle $\theta_B$, estimating $\omega_{dd}$ at each orientation. $\omega_{dd}$ oscillates over $\varphi_B$ with parameters encoding the spin's position. The azimuth of the spin $\varphi$ is encoded in the phase of the sine; the distance $r$, and polar angle $\theta$, may be extracted numerically by solving a set of nonlinear equations for the sine's offset and amplitude.

We illustrate a sweep of the magnetic field azimuth for two different target spin positions in Fig.\,\ref{fig:two_spins_example}, exhibiting distinct sine curves. To extract three variables, three sampling points are sufficient. However, utilizing the added information of the sinusoidal shape will provide a more robust estimation and validate the theory.

\section{Experimental results}

\begin{figure}
\includegraphics[scale=0.45]{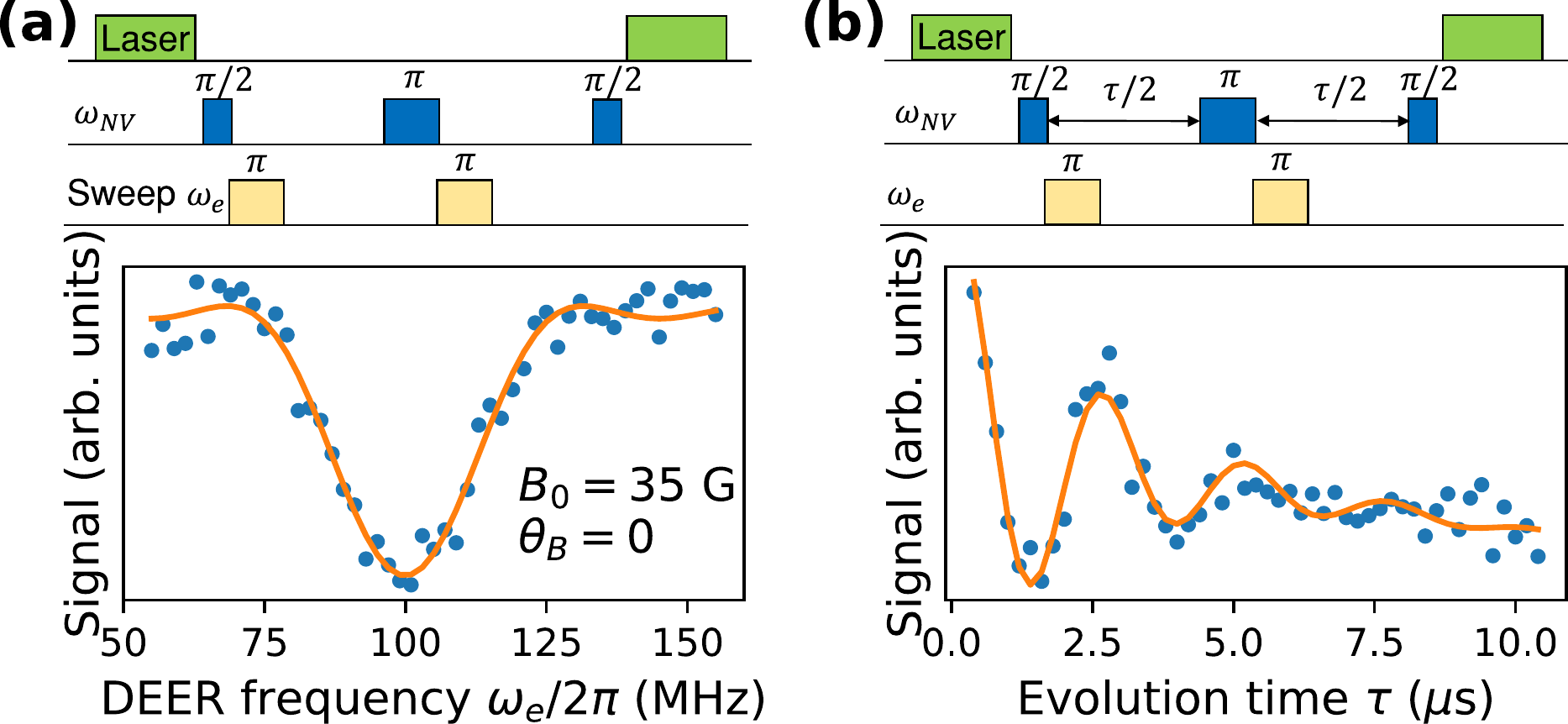}
\caption{\label{fig:deer_figure} \textbf{Measuring the dipolar coupling of proximate spins with DEER}. (a) The top panel depicts the DEER pulse sequence to measure the electron spin resonance of spins proximate to the sensor. The blue dots in the plot are the sensor signal as a function of the DEER frequency $\omega_e$, and the orange curve is a fit to squared sinc function. Here $B_0=35\ \mathrm{G}$ and aligned with the sensor's axis. (b)  The top panel shows the DEER pulse sequence to measure the dipolar coupling strength. We set $\omega_e$ to the resonance frequency (obtained in (a)), and the duration of the spin-echo $\tau$ was varied. The orange line is a fit to a decaying sine.}
\end{figure}

We demonstrate single spin mapping using magnetic tomography on a system composed of a shallow ($\sim8\,\mathrm{nm}$ depth) NV center in diamond as the sensor and a single proximate unpaired electron as the target spin. The target spin is possibly a stable surface spin~\cite{Grotz2011, Grinolds2014, Sushkov2014}, but its precise nature is unknown. We apply a constant magnetic field ($B_0$) with a permanent magnet, and control the magnitude and direction by moving the magnet.
We detect electron spins coupled to the sensor by a double electron-electron resonance (DEER) protocol, which has been discussed in the past in this context  \cite{Neumann2010b, Mamin2012a, Grinolds2014, Schlipf2017}. The pulse sequence (see Fig.\,\ref{fig:deer_figure}) consists of a spin-echo (Hahn) on the sensor spin, decoupling it from the surrounding spin-bath to extend the sensor's coherence \cite{Mamin2012a}. Flipping the target spin midway through the spin-echo couples the sensor spin's evolution with the field of the target spin. We acquire a resonance spectrum of surrounding spins by sweeping the target spin pulses' frequency $\omega_e$. Sweeping the duration ($\tau$) of the spin-echo evolution modulates the interaction time, and the resulting signal will be modulated according to the dipolar coupling strength of the spins \cite{Grinolds2014, Schlipf2017, Supplemental_Material}.

The sensor's spin evolution is usually affected by a large ensemble of spins (the bath), but due to the $r^{-3}$ scaling of the dipolar interaction, most of the signal originates from a volume of several nanometers radius around the sensor. If there are a few spins in this volume, their dipolar coupling frequencies will dominate, while the distant spin-bath will manifest as decoherence \cite{Grotz2011}. For a single target spin, the signal will oscillate at twice the dipolar coupling frequency (see derivation in the Supplemental Material \cite{Supplemental_Material}).

\begin{figure*}
\includegraphics[scale=0.59]{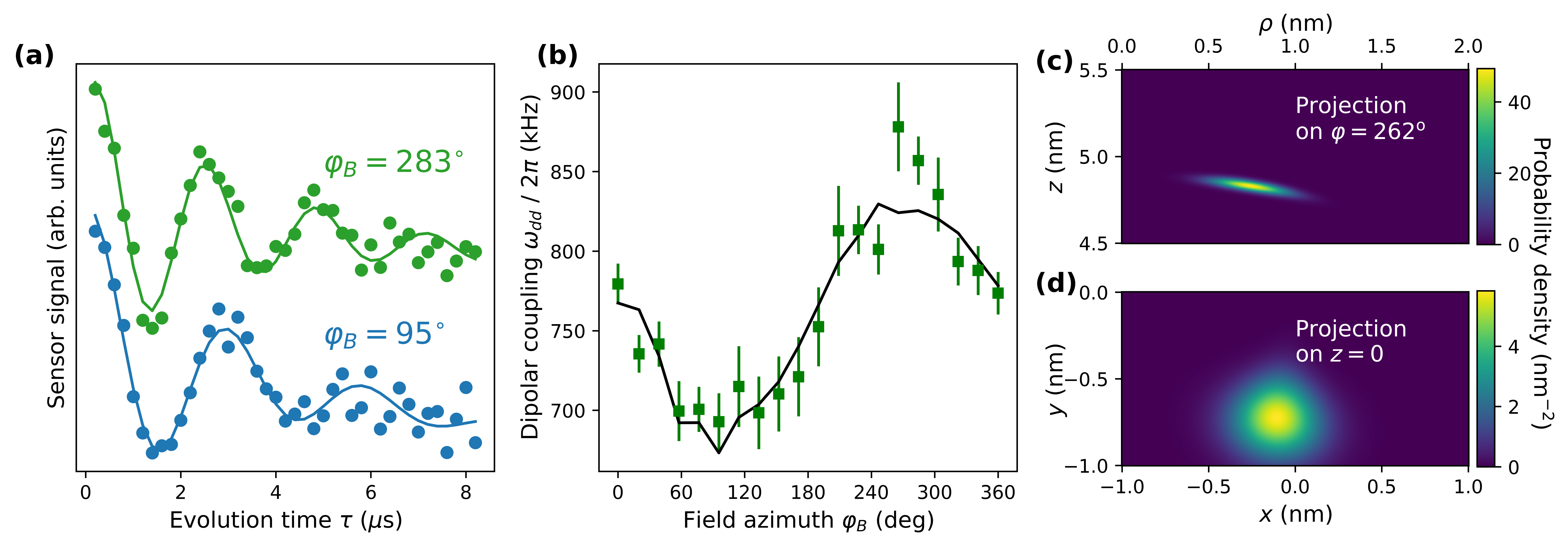}
\caption{\label{fig:experiment_results} \textbf{Locating a spin by magnetic tomography}. (a) Examples of dipolar coupling measurements for two different magnetic field azimuths $\varphi_B$, exhibiting two distinct dipolar coupling frequencies. Lines are fits to $f\left( \tau \right)=A\sin\left(\frac{1}{2}\omega_{dd}\tau+\phi\right)e^{-\frac{\tau}{T_2}}$. (b) The dipolar coupling frequency as a function of magnetic field azimuth $\varphi_B$, oscillating over a single period. The black line is a fit to Eq.\,\ref{wdd}, incorporating variations in $\theta_B$. We extract the target spin coordinates $\left(r,\theta,\varphi\right)$ from the fit. (c) and (d) Probability maps for the position of the target spin, as extracted from the data in figure (b), presented in a $\mathrm{\rho}$z-plot ($\rho\equiv\sqrt{x^2+y^2}$) and an xy-plot.}
\end{figure*}

To map the position of the proximate spin, we estimate the dipolar coupling strength at various orientations of the magnetic field. We measure at a low magnetic field ($B_0\approx38\ \mathrm{G}$) to minimize contrast loss due to a transverse magnetic field \cite{Tetienne2012}. Fig.\,\ref{fig:experiment_results} presents the results for a sweep of the magnetic field azimuth $\varphi_B \in \left[0, 360^{\circ} \right]$. In Fig.\,\ref{fig:experiment_results}(b), it is apparent that $\omega_{dd}\left(\varphi_B\right)$ oscillates over a single period, consistent with Eq.\,\ref{wdd}. The magnetic field was tilted at an average angle of $\left\langle\theta_B\right\rangle=19.4^{\circ}$, but the tilt angle $\theta_B$ varied over the sweep due to limited control of the magnetic field. We factored in $\theta_B$ variation by measuring the value of $\theta_B$ with the NV center for each data point. We then fitted the set of measured $\omega_{dd}\left(\varphi_B, \theta_B \right)$ to Eq.\,\ref{wdd} such that the fit function slightly deviates from a smooth sine shape as expected if $\omega_{dd}$ was a function of only $\varphi_B$ (see Supplemental Material for details on data fitting \cite{Supplemental_Material}). From the fit, we obtain the target spin coordinates:
\begin{eqnarray}
r=4.89\pm0.02\,\mathrm{nm};\,\theta=9.0\pm0.9^{\circ};\,\varphi=-98\pm6^{\circ}
\label{eq:extracted_coords}
\end{eqnarray}

We repeated the experiment for several trajectories of the magnet, corresponding to different values of magnetic field tilt $\theta_B$. The $\omega_{dd}$ modulation gradually increased for larger field tilts, consistent with Eq.\,\ref{wdd} (see Supplemental Material for data \cite{Supplemental_Material}).

\section{Spin location precision}

To quantify the precision of the measurement, we define a location uncertainty based on the coordinate uncertainties obtained when fitting data to Eq.\,\ref{wdd}:
\begin{eqnarray}
&& \Delta R \equiv \ \left( 8r^2 \sin\left( \theta \right) \Delta r \Delta \theta \Delta \varphi \right)^{\frac{1}{3}}
\label{eq:Delta_V}
\end{eqnarray}
\noindent
where $\Delta x_i$ is the interval of confidence for coordinate $x_i$.\\
From the fit of the experimental data (Fig.\,\ref{fig:experiment_results}(b)) we estimate the uncertainty of the target spin position to be $\Delta R = 0.09\,\mathrm{nm} $. Fig.\,\ref{fig:experiment_results}(c) depicts the target spin position probability map. The largest uncertainty is along the $\hat{\varphi}$ axis, with $r \sin \left(\theta\right) \Delta \varphi=0.09\ \mathrm{nm}$.

\begin{figure}
\includegraphics[scale=0.5]{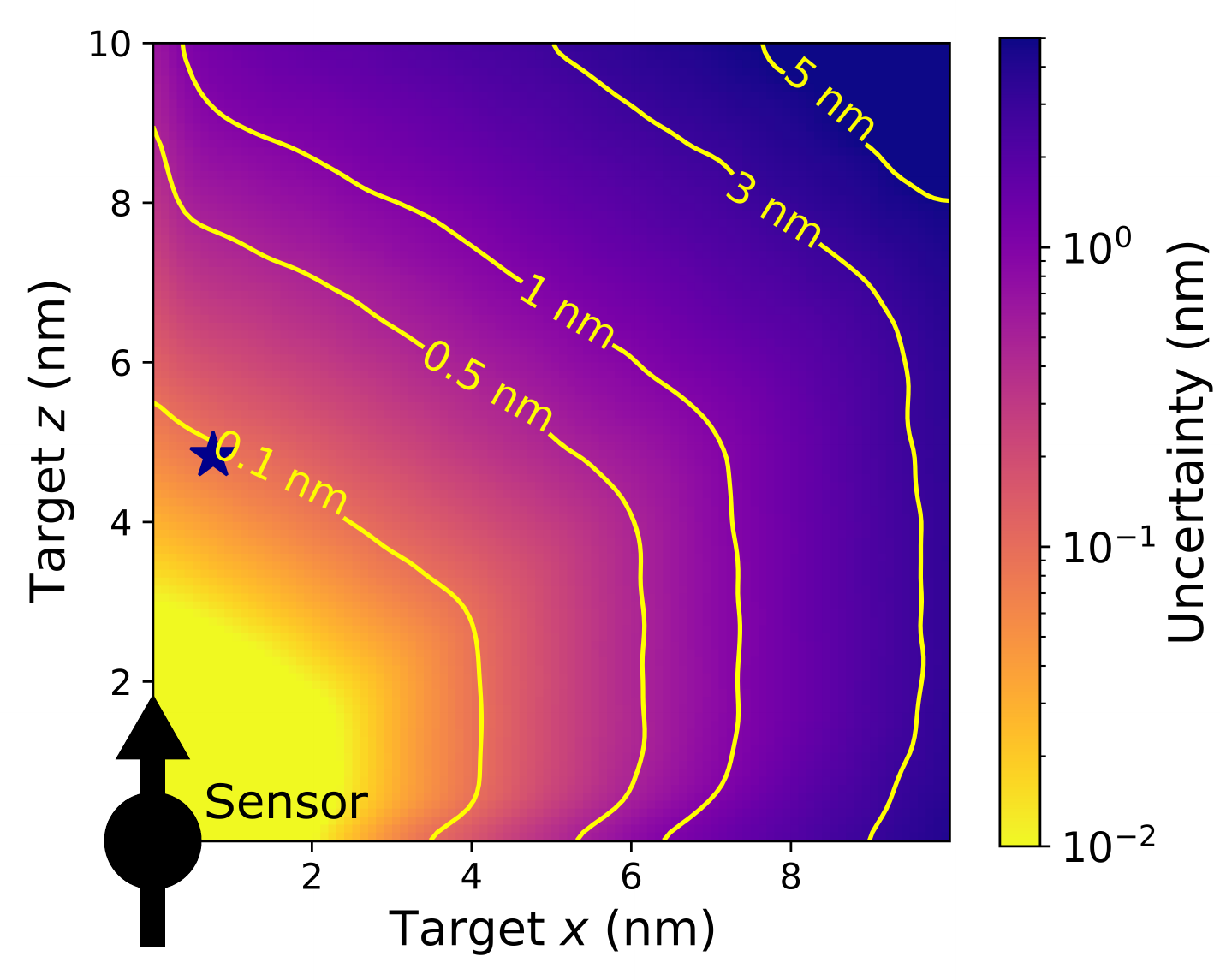}
\caption{\label{fig:uncertainty} \textbf{Location precision of electron spins.} The location uncertainty $ \Delta R $ vs. target spin coordinates ($x=r\sin\theta,\ z=r\cos\theta$), assuming $\Delta \omega_{dd} = 20\ \mathrm{kHz}$. Contours denote equal levels of uncertainty. The blue star marks the experiment's spin position.}
\end{figure}

To study the relevance of the method, we study the dependence of the location precision $\Delta R$ on the spin's position and sensor's characteristics. Underlying $\Delta R$ is the dipolar frequency sensitivity, which depends on the system parameters (e.g., the sensor decoherence time $T_2$), and the specific protocol \cite{Degen2017}. To focus the discussion on unique aspects of this measurement, we assumed a given frequency uncertainty $\Delta \omega_{dd}$ and calculated the uncertainty’s dependence on the position of a target spin. The location uncertainty is proportional to the frequency uncertainty ($\Delta R \propto \Delta \omega_{dd}$), so the functional dependence of $\Delta R \left( \mathbf{r} \right)$ is independent of the choice of $\Delta \omega_{dd}$. For the discussion, we use $\Delta \omega_{dd}=2\pi\times 20\,\mathrm{kHz}$ (in our experiment the uncertainty was in the range of 12-28 kHz), and calculated $\Delta R \left( r,\,\theta \right)$, plotted in Fig.\,\ref{fig:uncertainty}. Due to the symmetry of $\omega_{dd}(\mathbf{r}, \mathbf{\hat{b}})$, $\Delta R$ does not depend on the target spin's azimuth $\varphi$.

We find that the uncertainty is minimal for $\theta \rightarrow 0$, and maximal for $\theta \rightarrow 90^\circ$. At $\theta=90^\circ $, the second term of Eq.\,\ref{wdd} vanishes, eliminating the information on the spin's azimuth $\varphi$, such that for spins near $\theta=90^\circ $, we can infer only the distance $r$ and polar angle $\theta$. Also, Eq.\,\ref{wdd} is centrosymmetric, so there will always be (at least) two solutions for every data set. In many scenarios, however, these ambiguities may be resolved with prior information on the system. For example, when imaging a sample on the crystal's surface, target spins will be in a thin slice on the surface, within a single hemisphere around the sensor \cite{Staudacher2013}. Nonetheless, target spins over 10\,nm away from the sensor in the range of $ \theta \in \left[ 0, 45^\circ \right]$ may be located with a precision better than 2 nm. This fact makes it appealing to use sensors whose axis is normal to the surface, e.g., an NV center in a diamond crystal cut along the $\left( 111 \right)$ crystal plane \cite{Michl2014PerfectSurfaces}.

The protocol is based on a tilted magnetic field, and we also explore the impact of the field parameters -- magnitude $B_0$, and tilt $\theta_B$ -- for the case of an NV center sensor. The NV center's function as a magnetometer usually relies on an optical measurement of the spin state within a two-state subspace of the spin-1 states, e.g. $\vert m_s=0\rangle$, $\vert m_s=+1\rangle$. The optical contrast between the states is maximal for a field aligned with the center's zero-field splitting axis, and decreases in the presence of a transverse field \cite{Tetienne2012}. As the measurement requires a transverse magnetic field to modulate the dipolar coupling, there is a competition between the modulation amplitude, and the contrast drop. To find the optimal conditions, we calculated the uncertainty as a function of magnetic field $B_0$ and field tilt angle $\theta_B$ and incorporated the optical contrast. At fields near $500\,\mathrm{G}$, there is a sharp decrease in contrast for any finite transverse field, and the uncertainty is large. For any field satisfying the condition $ \mathcal{H}_{dd} \ll\omega_0\ll D $, the dipolar coupling modulation amplitude does not depend on $B_0$, and only on $\theta_B$. The criterion is satisfied for $B_0 \gtrsim 10\,\mathrm{G}$, and so we calculated the expected uncertainty by solving the Hamiltonian of the system. We find that the uncertainty is minimized for a maximal tilt of $\theta_B \approx 85^{\circ}$ and minimal field $B_0\sim10\,\mathrm{G}$ (see Supplemental Material for details \cite{Supplemental_Material}).

\section{Applicability to nitroxide spin-labels}

So far, we have discussed locating an electron spin that does not interact with nuclear spins. However, in many scenarios, the target electron spin may have significant interactions with nearby nuclear spins, as is the case with nitroxide radicals, the most common type of spin-labels. Adding a hyperfine interaction term with a nucleus, $\mathbf{S}_e \cdot \mathbb{A}_{hf} \cdot \mathbf{I}$, modifies the spin Hamiltonian (Eq.\,\ref{eq:spin_hamiltonian}). We expect a subsequent modification to $\omega_{dd}\left( \mathbf{r}, \mathbf{\hat{b}} \right)$ (Eq.\,\ref{wdd}), as the assumption that the Zeeman term dominates the eigenstates of the target spin is no longer valid.

For this discussion, we focus on the case of nitroxide radicals, where the electron spin is coupled by hyperfine constants of $\sim 2\pi \times 100\,\mathrm{MHz}$ to the adjacent nitrogen nuclear spin (\textsuperscript{14}N or \textsuperscript{15}N)~\cite{Marsh2019Spin-LabelSpectroscopy}. At fields of tens of Gauss, the hyperfine term is comparable to $\gamma_e B_0$. We study the applicability of the magnetic tomography method to nitroxide spin-labels by numerically simulating a magnetic azimuth sweep of the dipolar coupling between an NV sensor and a nitroxide spin-label. We use typical hyperfine coupling parameters for a nitroxide radical with an \textsuperscript{14}N nuclear spin ($A_\parallel = 2\pi\times 101.4\,\mathrm{MHz}, A_\perp = 2\pi\times 14.7\,\mathrm{MHz}$) \cite{Marsh2019Spin-LabelSpectroscopy}.

\begin{figure}
\includegraphics[scale=0.54]{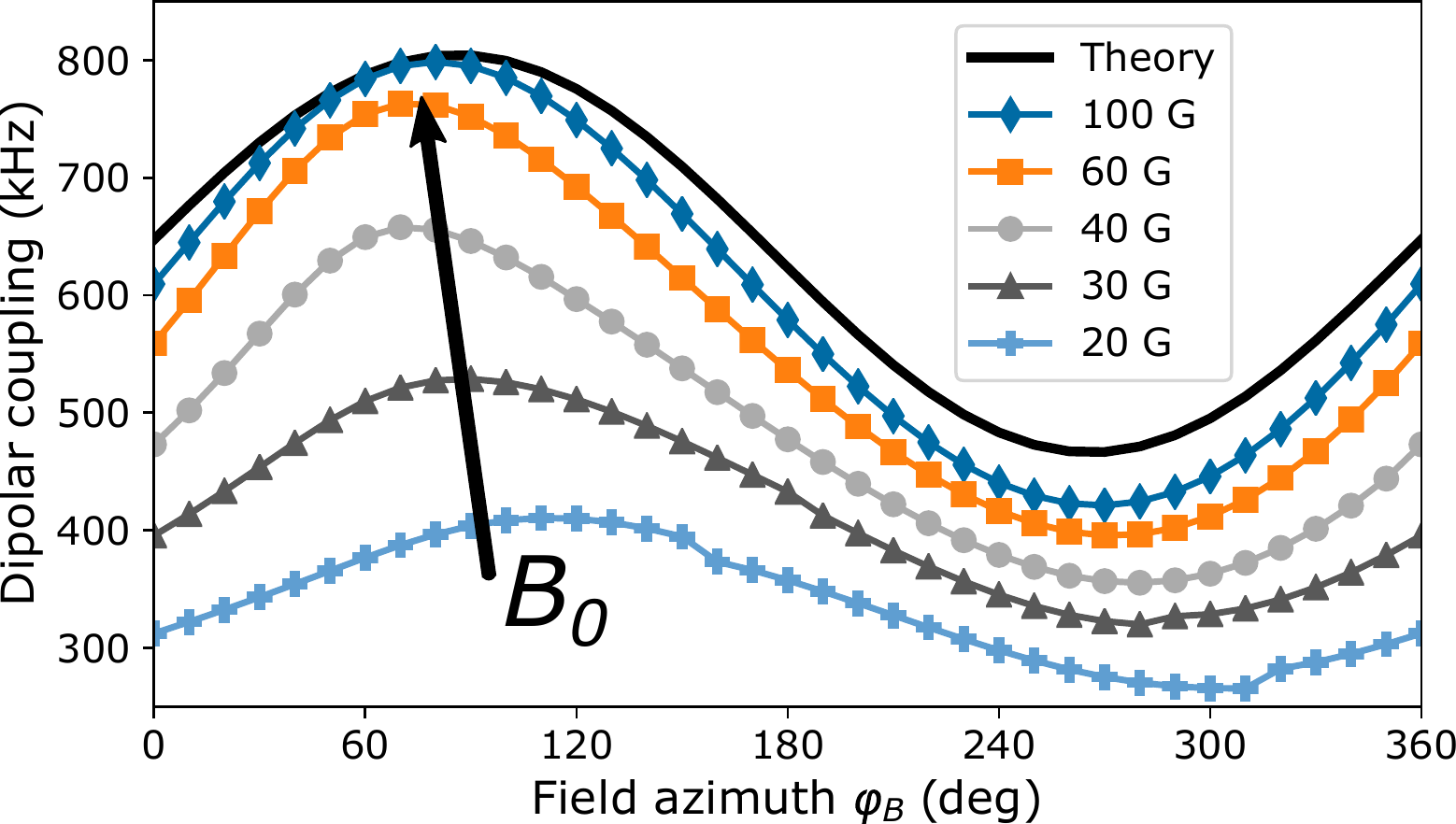}
\caption{\label{fig:nitroxide} \textbf{Magnetic azimuth sweep of nitroxide spin labels.} A comparison of numerically calculated azimuth sweeps of a nitroxide radical (\textsuperscript{14}N) at different field magnitudes. The electron spin is arbitrarily positioned, with an arbitrary orientation of the hyperfine axes. The curves are compared to the theoretical model for $\omega_{dd}\left(\mathbf{r}, \mathbf{\hat{b}} \right)$. At $B_0\gtrsim100\,\mathrm{G}$, the calculated curve is well approximated by the theoretical model.}
\end{figure}

We calculate the modulation of the dipolar coupling with a spin-label at an arbitrary position and orientation, under $B_0$ fields in the range of 20-100\,G. We assume that the spin-label is in a thermal state of the nuclear spin, i.e., equal probabilities of the nucleus spin-1 states. The simulations are presented in Fig.\,\ref{fig:nitroxide}, and compared with the theoretical model for $\omega_{dd}\left(\mathbf{r}, \mathbf{\hat{b}} \right)$ (Eq.\,\ref{wdd}). At 20\,G, the simulated azimuth sweep deviates significantly from the model and approaches it gradually for higher fields. At a field of $B_0 = 100\,\mathrm{G}$ the simulated curve is well approximated by the theoretical model. While a field of 100\,G entails a decrease in precision due to the transverse field, our calculations show that it allows mapping spins with \AA ngstrom-precision\,\cite{Supplemental_Material}. For the case of a hyperfine coupled spin, however, a more precise location estimation may be obtained by fitting to a calculated $\omega_{dd}\left( \mathbf{r}, \mathbf{B} \right)$, instead of the analytical term.

The magnetic tomography model $\omega_{dd}\left(\mathbf{r}, \mathbf{\hat{b}} \right)$ of Eq.\,\ref{wdd} applies for a hyperfine coupled spin already at a moderate field of 100\,G despite the Zeeman term being on the same order of the hyperfine parameter ($A_{\parallel}\approx 2\pi \times 101.4\,\mathrm{MHz} \sim \gamma_e B_0 \approx 2\pi \times 280\,\mathrm{MHz}$), and so it is not negligible. This stems from the fact that, for a thermal ensemble of the nuclear spin states at 100\,G, the expectation value of the target electron spin operator perpendicular to the field axis $\mathbf{\hat{b}}$ satisfies $\left| \braket{S_e^{\perp}} / \braket{S_e^{b}} \right| \ll 1$ \cite{Supplemental_Material}. It follows that the secular approximation holds for this scenario as well, and so Eq.\,\ref{wdd} constitutes a valid approximation. Thus, the magnetic tomography method may also be applicable to mapping nitroxide spin-labels, and other similar hyperfine-coupled electron spins.

\section{Discussion and Conclusions}

NV centers in diamond are a leading platform for nanoscale magnetometry, particularly for single-molecule magnetic resonance tasks. Here, we demonstrated a method to map the locations of spins in the vicinity of an NV center sensor with {\AA}ngstrom-scale precision. The location precision of the spin we demonstrated here is one order of magnitude higher than previously reported for a similar magnetic field scanning experiment ($\mathrm{\sim 1\, nm}$) \cite{Sushkov2014} and a spin imaging technique based on a scanning magnetic tip ($\mathrm{1.5\,nm}$) \cite{Grinolds2014}. Magnetic resonance imaging demonstrated recently with a scanning tunneling microscope exhibited superior precision but requires strict conditions \cite{Willke2019, Willke2021CoherentSurface}. Notably, the magnetic tomography method does not require a scanning probe setup, which is operationally complex, and the method is operable at both ambient and cryogenic conditions.

Spin mapping with {\AA}ngstrom resolution may provide added value for applications such as single-molecule distance measurements. For this, we would measure the positions of a pair of spin-labels attached to a biomolecule, from which we can infer the distance. To do so, the dipolar coupling to each spin needs to be measured as a function of the magnetic field direction, and fitted to $\omega_{dd}\left( \mathbf{r}, \mathbf{\hat{b}} \right)$ (Eq.\,\ref{wdd}). A minimal frequency resolution is needed to distinguish between the dipolar coupling of two or more spins, and it is given by the coherence time of the sensor $\delta \omega \sim \frac{1}{T_2}$. Another approach is selective addressing by separating the targets' electron spin resonances, which allows measuring each spin's dipolar coupling separately~\cite{Cooper2020IdentificationDiamond}. Selective addressing may be achieved also by attaching spin-labels with distinct resonance spectra, such as nitroxide radicals with different nitrogen isotopes (\textsuperscript{14}N, \textsuperscript{15}N) \cite{Lee1984, MunueraJavaloy}.

The uncertainty of the measurement is proportional to the frequency estimation uncertainty, so techniques that lengthen the coherence time and improve the sensor readout efficiency would enhance the spin location precision. Nonetheless, with nanometer precision up to 10 nm away from the sensor, the method can be used for sensing spin-labels on molecules external to the diamond crystal. Thus, the magnetic tomography method is relevant for studying the structure of individual molecules by spin-label distance measurements, or high-resolution characterization of quantum spin networks.

\begin{acknowledgments}
We acknowledge Leora Schein-Lubomirsky for contributions to the experimental setup, and calculations. We thank Alon Salhov for ideas and fruitful comments. It is a pleasure to thank Durga Dasari for insightful discussions. A.\,F.\,is the incumbent of the Elaine Blond Career Development Chair and acknowledges the historic generosity of the Harold Perlman Family, research grants from the Abramson Family Center for Young Scientists and the Willner Family Leadership Institute for the Weizmann Institute of Science, as well as support from the Israel Science Foundation (ISF 963/19, ISF 419/20).
\end{acknowledgments}

\nocite{Sangtawesin2019a, Binder2017, Aharonovich2011Diamond-basedEmitters, schweiger2001principles, Tetienne2012, Maertz2010a, Degen2017, Marsh2019Spin-LabelSpectroscopy}
\bibliography{mapping}

\end{document}